\documentclass[conference]{IEEEtran}
\IEEEoverridecommandlockouts
\usepackage{cite}
\usepackage{amsmath,amssymb,amsfonts}
\usepackage{algorithmic}
\usepackage{graphicx}
\usepackage{textcomp}
\usepackage{xcolor}
\usepackage{lipsum}
\usepackage{graphicx}
\def\BibTeX{{\rm B\kern-.05em{\sc i\kern-.025em b}\kern-.08em
    T\kern-.1667em\lower.7ex\hbox{E}\kern-.125emX}}
\begin{document}

\title{ICS-CTM2: Industrial Control System Cybersecurity Testbed Maturity Model\\
}

\author{\IEEEauthorblockN{Souradeep Bhattacharya, Burhan Hyder, Manimaran Govindarasu }
\IEEEauthorblockA{\textit{Department of Electrical and Computer Engineering} \\
\textit{Iowa State University, Ames, IA 50011}\\
Email: sbhatta@iastate.edu, bhyder@iastate.edu, gmani@iastate.du}
}

\maketitle

\begin{abstract}
Industrial Control System (ICS) testbeds serve as a platform for evaluating and validating control system performances, cybersecurity tools and technologies. In order to build or enhance an ICS testbed, it is vital to have a deeper understanding of its design specifications and characteristic attributes. Satisfying this prerequisite involves examination and assessment of these attributes for existing testbeds. To further increase confidence in a testbed's functionality, it is important to perform a comparative analysis of its specifications with other ICS testbeds.  
However, at present, there is no standardized methodology available to provide a comparative assessment of different testbeds. In this paper, we propose a methodology for analyzing ICS testbeds, inspired by the Cybersecurity Capability Maturity Model (C2M2). In particular, we then define a ICS Cybersecurity Testbed Maturity Model, its domains, and the associated maturity indicator levels. To demonstrate the benefit of the model, we have conducted a case study analysis for several ICS testbeds, representing different industrial sectors. Our analysis provides deeper insights into the relative strengths and limitations of these testbeds, together with scope for future enhancements, with respect to the domains defined by the model.
\end{abstract}

\vspace{-0.2cm}
\section{Introduction}
In recent years, with the emergence of Industry 4.0 criterion, there has been a considerable increase in digitization of legacy Industrial Control Systems (ICS). This accelerated process has resulted in growing convergence between information technologies (IT) and operational technologies (OT), thereby increasing the inter-connectivity of different systems. Owing to advancement in communication, information, and operational technologies, new approaches in the industrial chain have risen enabling tremendous growth for ICS, such as industrial internet of things, big data analytics, cloud computing, robotics and human-machine interactions \cite{Alcaraz2019}. However, the paradigm shift towards digitization and inter-connectivity exposes ICSs to new vulnerabilities and challenges. This in turn increases the attack surfaces for adversaries to exploit. Therefore, academia, industry and government entities are increasingly proposing development of ICS testbeds for evaluating tools and technologies designed to assess system performance metrics and enhance cybersecurity capabilities. 

ICSs interconnect, monitor, and control processes in a variety of industries, such as electric power generation and distribution, transportation, water, gas, oil, chemical production, and manufacturing. ICS consist of two major components - the OT network, and the IT network. The OT network comprises of hardware and software that are used to monitor and manage industrial equipment, processes, assets, and events, e.g., Programmable Logic Controllers (PLC), Supervisory Control and Data Acquisition (SCADA), Distributed Control System (DCS), sensors and actuators. The IT network comprises of workstations, servers, databases, routers and network switches, which are used to control and manipulate the flow of information.

This paper is structured as follows. In Section-I, we provide a brief introduction of ICS, highlighting the differences between IT and OT networks. In Section-II, we discuss why it is crucial to develop testbeds for ICS and what are the requirements to consider. We also refer some of the related works in this field. Lastly, in this section we have also discussed the need for developing a systematic testbed assessment methodology for analyzing different ICS testbeds. In Section-III we describes the design and implementation methodology for the proposed ICS Cybersecurity Testbed Maturity Model (ICS-CTM2) framework. Finally, Section-IV provides a case study for ICS testbed analysis using the proposed ICS-CTM2 framework, and also presents a visualization of the results.

\section{Overview of ICS Testbeds}
A testbed can be defined as a testing environment or a platform that simulates real-world activities under supervision, in order to conduct experimental evaluations. When researchers need to work with real-world ICS environments for examining new technologies, testbeds provide a validated ecosystem representing the real-world process.

\subsection{Why Do We Need A Testbed For ICS}
Testing IT security systems is a mature, well-defined field with a plethora of best practices, procedures and information. However, testing industrial control systems and OT networks is a comparatively newer frontier \cite{Khorrami2016}. Evolution of today's ICS has been a causal effect of amalgamation of IT capabilities into legacy physical systems, often replacing or augmenting existing physical control operations. However, it also comes with various challenges that need to be mitigated - cybersecurity being critical. As per annual reports published by Kaspersky, it can be observed that there is a considerable increase in vulnerabilities detected in ICS every year (19 in 2010, 189 in 2015, 415 in 2018, and 509 in 2019) \cite{Kaspersky2020}. Therefore, developing custom testing environments imitating real-world solutions are a crucial component in ICS research. In addition to \textit{vulnerability research}, cyber-physical ICS testbeds provide several applications such as \textit{impact analysis} (capturing risk posed by security events); \textit{mitigation evaluation} (to increase robustness of cyber infrastructure); \textit{cyber-physical metrics} (functional tests for system reliability and performance); \textit{data and model development} (models and datasets to facilitate more accurate analysis); \textit{security validation} (assessment of cybersecurity compliance requirements); \textit{cyber forensics} (analysis of cyber attacks); \textit{operator training} and \textit{education} \cite{Hahn2013}.

\subsection{Requirements For Developing An ICS Testbed}
Building an ICS testbed is a cost-dependent, resource-intensive, and complex task, involving collaboration of multiple disciplines. As it is essential to perform experimental activities and generate data from different industrial scenarios, it is imperative to develop custom testbeds accurately replicating the respective industrial practices.
\subsubsection{Testbed Classification}Depending on the process, protocol, and infrastructure elements involved in a testbed, it can be classified into one of the four categories - \textit{Physical}, \textit{Simulation}, \textit{Virtual}, or \textit{Hybrid} \cite{Holm2015}. In a physical testbed, real hardware and software are deployed in both the cyber and physical layers. Although, it aims to be as close to the real system as possible providing highest fidelity, physical testbeds are usually expensive and have a long development cycle. Simulation testbeds are based on software simulations, whereas virtual testbeds involve both simulations and emulation of hardware components. Simulation and virtual testbeds provide a low-cost alternative to physical testbeds, and are more flexible to upgrades. However, these testbeds may lack precision and often does not provide real-time operation. The most commonly used approach is the hybrid testbed. These testbeds are designed through a combination of physical devices, simulation, and virtualized components. Hybrid testbeds provides lower fidelity than physical testbeds, but allows for lower development cost and higher scalability.
\subsubsection{Testbed Properties}Any practical testbed is required to satisfy certain clearly defined objectives. These objectives constitute the properties of an ICS testbed which indicate its reliability \cite{Ani2021}. Although, it may be challenging to meet every objective, it is vital for researchers to ascertain an effective and optimized trade-off criteria during the design phase. All properties pertaining to an ICS testbed can be grouped under three major inter-related properties as follows:

\textbf{Fidelity:} Fidelity of a testbed can be defined as how closely and accurately the testbed replicates a real-world ICS \cite{Ani2021}, \cite{Conti2021}. The properties which directly impact fidelity are: \textit{Measurability} (allowing assessment of experiments), \textit{Measurement Accuracy} (degree of correctness of experimental results), \textit{Repeatability} (ability to replicate experiments), \textit{Reproducibility} (ability to produce consistent results), \textit{Usability} (ability to be utilized for defined purposes), \textit{Complexity} (transparency of architecture), and \textit{Safety} (safe operation of testbed).

\textbf{Scale:} Scalability of a testbed can be defined as a testbed's ability to be expanded in functionality, architecture, and, infrastructure \cite{Hahn2013}, \cite{Ani2021}. The properties which directly impact scale are: \textit{Cyber-Physical Integration}, \textit{Reconfigurability} (ability to accommodate new design or components), \textit{Extensibility} (ability to broaden functional scope), \textit{Adaptability} (ability to support new or alternative test cases), \textit{Interoperability} (ability to support any combination of simulation or hybrid approaches), \textit{Federation} (inter-connectivity of multiple testbeds), and \textit{Critical Infrastructure}.

\textbf{Cost:} Cost of a testbed depends on the approximate implementation cost of the process, hardware, software, and licenses \cite{Hahn2013}, \cite{Conti2021}. The properties which directly impact cost are: \textit{Diversity} (ability to utilize products and services from multiple vendors), \textit{Openness} (ability to support open-source products and remote access), \textit{Training \& Development} (for education of operators), and \textit{Classification} (Physical, Simulation, Virtual, or Hybrid).

\subsubsection{Compliance of Industrial Standards}NIST SP 800-82 Rev.2 \cite{Stouffer2015} provides a guideline for ICS Security. It specifies that the basic architecture for an ICS environment must include four core components: \textit{Physical Process}, \textit{Field Devices}, \textit{Communication Architecture}, and \textit{Control Center}. Therefore, adherence to standardized design attributes contribute significantly to the reliability and trustworthiness of ICS testbeds. 

\subsection{Related Work}
In this paper, we directed our efforts towards performing an extensive review of the current state in the field of Industrial Control System (ICS) testbeds. Among the existing literature, survey article \cite{Holm2015} focuses on exploring the primary objectives and application possibilities of 30 different ICS testbeds. Another survey \cite{Conti2021} provides a comprehensive classification of ICS testbeds into three groups (physical, virtual, or hybrid), along with an analysis of ICS datasets available for developing new IDS security mechanisms. \cite{Christiansson2008} provides an outline for developing a SCADA Security Testbed. Some existing literature (e.g. \cite{Ashok2016}, \cite{Kovalenko2017}) provides detailed description of testbed architecture and design, implementation and experiments conducted. For instance, \cite{Ashok2016} discusses the PowerCyber testbed at Iowa State University, which models a smart electric grid, and \cite{Kovalenko2017} discusses the SMART testbed at University of Michigan, which models a warehouse production line.

\begin{figure*}
    \centering
    \includegraphics[width=\textwidth]{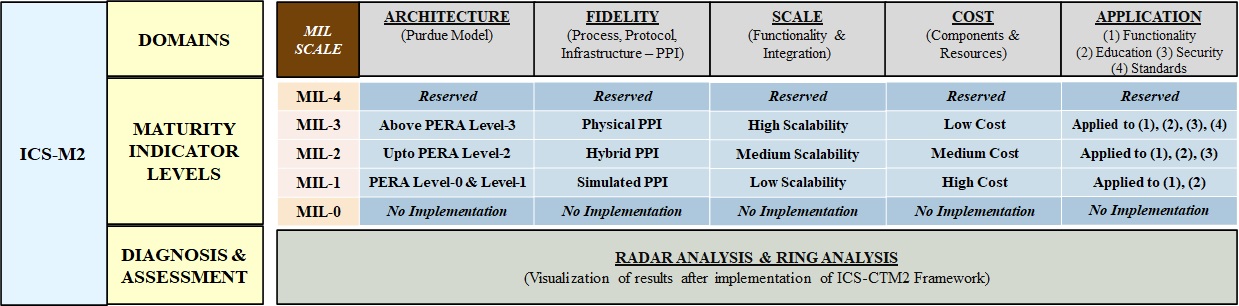}
    \caption{The Proposed ICS Cybersecurity Testbed Maturity Model (ICS-CTM2) Framework}
    \label{fig:ICS-M2}
    \vspace{-0.5cm}
\end{figure*}

\subsection{Assessment of ICS Testbeds}
Although survey articles provide a compilation of different testbeds, they lack any quantitative assessment methodology which makes it difficult to easily recognise relevant variations in the listed testbeds. Literature pertaining to design and implementation of ICS testbeds focus only on the individual testbeds, without any comparative analysis with other testbeds. To the best of our knowledge, there is no standardized methodology available at present, to provide a comparative assessment of different testbeds. Therefore, a systematic methodology is needed which can provide a focal guide for researchers developing future ICS testbeds or enhancing current testbeds.

\section{Proposed ICS Cybersecurity Testbed Maturity Model}
The central objective of the ICS Cybersecurity Testbed Maturity Model (ICS-CTM2) framework is to provide a standardized and systematic method for comprehensive and comparative analysis of ICS testbed design specifications. The ICS-CTM2 framework is inspired from the Energy Subsector-Cybersecurity Capability Maturity Model (ES-C2M2) developed by the US energy sector and Department of Energy (DOE) \cite{C2M22014}.

Figure 1 shows the conceptual diagram for the proposed ICS-CTM2 framework. Table I provides a list of ICS testbeds that we have considered for the case study analysis of ICS-CTM2 and is further described in detail under Section IV.

\textbf{ES-C2M2:} The ES-C2M2 \cite{C2M22014} is a tool which allows electric utilities and grid operators to assess, evaluate, and improve their cybersecurity capabilities, and also to prioritize their actions and investments for enhancing cybersecurity. The ES-C2M2 was developed in 2012 as part of a White House initiative led by the Department of Energy (DOE) in partnership with the Department of Homeland Security (DHS), involving collaboration with industry members, other Federal agencies and stakeholders. ES-C2M2 provides a toolkit (available on request from DOE) that can be utilized by any organization to measure and improve its cybersecurity program. Owing to its self-evaluation methodology and industry focused application which makes ES-C2M2 an easily scalable tool, in this paper we adapted ES-C2M2 to develop our ICS-CTM2 framework.

\subsection{Methodology}\label{AA}
Maturity model can be defined as an organized and structured method to convey a path of knowledge, experience, and learning. Such assessment models enable researchers and organizations to evaluate and make improvements to their ICS testbeds, by providing standardized benchmarks. Inspired by ES-C2M2, the ICS-CTM2 model arises from a combination of existing industrial standards, frameworks, and procedures adapted for developing an ICS testbed. The ICS-CTM2 architecture comprises of the following three sections, also illustrated in Figure-1:
\subsubsection{Domains}
ICS-CTM2 domains represent a structured set of parameters that include essential information related to an ICS testbed. ICS-CTM2 framework consists of five domains (\textit{Architecture}, \textit{Fidelity}, \textit{Scale}, \textit{Cost}, and \textit{Application}) based on which each testbed can be evaluated and its capabilities can be determined. For each domain, the model provides a description, which summarizes the objective of the domain along with the evaluation criterion considered within the context of the domain. The defined objective is then assessed by determining the extent of implementation of each evaluation criteria.

Figure 2 describes the evaluation criteria for each domain in ICS-CTM2 framework. Based on this assessment, a maturity level is assigned to each domain for a testbed. The five domains for ICS-CTM2 model are:

\begin{table}[ht]
\caption{list of selected testbeds for ICS-CTM2 case study analysis}
  \vspace{-0.2cm}
\begin{tabular}{|c|c|c|c|c|c|}
\hline
\multicolumn{1}{|c|}{{\textbf{Sl.No.}}} & \multicolumn{1}{|c|}{{\textbf{Institute}}} & \multicolumn{1}{|c|}{{\textbf{Industry}}} \\ \hline
\multicolumn{1}{|c|}{{1}} & \multicolumn{1}{l|}{{Iowa State University (ISU) \cite{Ashok2016}}} & \multicolumn{1}{l|}{{\textit{Smart Grid}}}\\ \hline
\multicolumn{1}{|c|}{{2}} & \multicolumn{1}{l|}{{SUTD, Singapore \cite{Mathur2016}}} & \multicolumn{1}{l|}{{\textit{Water Treatment}}} \\ \hline
\multicolumn{1}{|c|}{{3}} & \multicolumn{1}{l|}{{University of Michigan (U-M) \cite{Kovalenko2017}}} & \multicolumn{1}{l|}{{\textit{Manufacturing}}} \\ \hline
\multicolumn{1}{|c|}{{4}} & \multicolumn{1}{l|}{{ORNL \cite{Gillen2020}}} & \multicolumn{1}{l|}{{\textit{HVAC/Cooling}}} \\ \hline
\multicolumn{1}{|c|}{{5}} & \multicolumn{1}{l|}{{Univ. of Alabama, Huntsville (UAH) \cite{Alves2016}}} & \multicolumn{1}{l|}{{\textit{Gas Pipeline}}} \\ \hline
\multicolumn{1}{|c|}{{6}} & \multicolumn{1}{l|}{{Univ. of Tennessee (U.T) \cite{Zhang2019}}} & \multicolumn{1}{l|}{{\textit{Nuclear Plant}}} \\ \hline
\multicolumn{1}{|c|}{{7}} & \multicolumn{1}{l|}{{Ohio State University (OSU) \cite{Oruganti2019}}} & \multicolumn{1}{l|}{{\textit{Automobile}}} \\ \hline
\multicolumn{1}{|c|}{{8}} & \multicolumn{1}{l|}{NIST \cite{Candell2014}} & \multicolumn{1}{l|}{{\textit{Multi-scenario}}} \\ \hline
\end{tabular}
\end{table}

\begin{figure*}
    \centering
    \includegraphics[width=\textwidth]{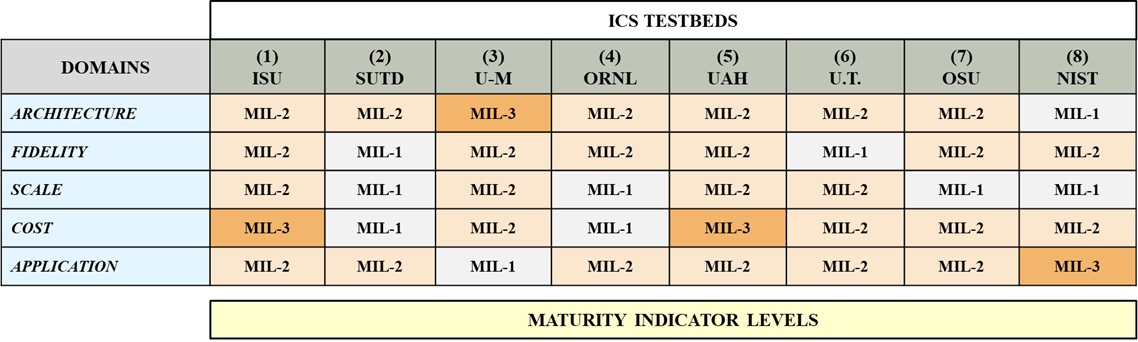}
    \caption{Case Study: Maturity Indicator Levels of each ICS-CTM2 domain for selected ICS Testbeds.}
    \label{fig:ICS-M2}
    \vspace{-0.5cm}
\end{figure*}

\begin{figure}[h]
    \centering
    \includegraphics[width=8.5cm]{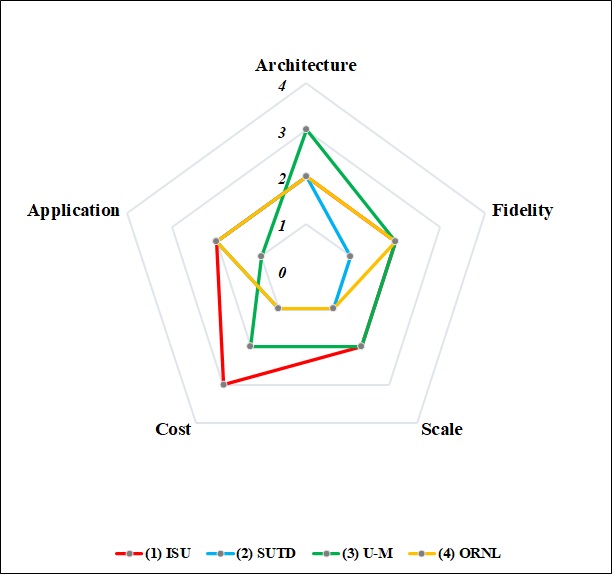}
    \caption{Case Study: Radar Analysis for ICS Testbeds (1)-(4)}
    \label{fig:Radar}
    \vspace{-0.5cm}
\end{figure}

\begin{figure}[h]
    \centering
    \includegraphics[width=8.5cm]{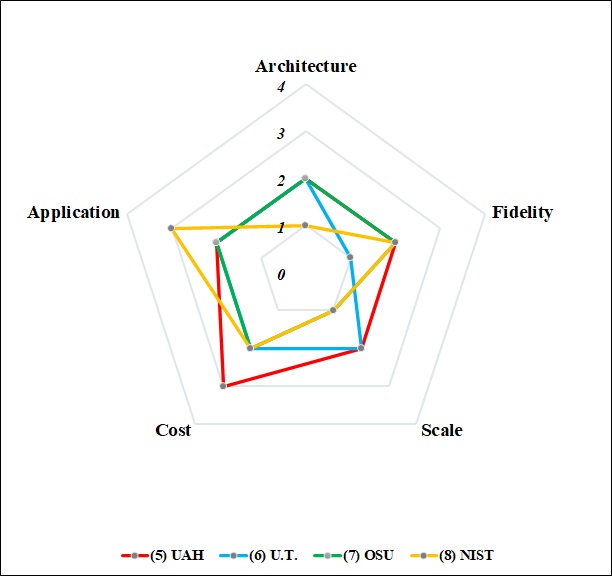}
    \caption{Case Study: Radar Analysis for ICS Testbeds (5)-(8)}
    \label{fig:Radar}
    \vspace{-0.5cm}
\end{figure}

\textbf{Architecture:} The Purdue Enterprise Reference Architecture (PERA) \cite{Williams1994} is the most widely accepted and utilized ICS network architecture model in the industry. Since ICS consists of different equipments and protocols across different industries and regions, PERA serves as a common reference model for evaluating this domain of ICS-CTM2 framework. PERA was developed by Theodore J. Williams in 1993, as a collaborative effort between Purdue University and members of the industry. This model has been adopted by several industry security standards such as ISA-99 (ISA/IEC 62443) \cite{ISA2007} and is used as the foundation for ICS IT/OT network design. PERA is segmented into five levels as described below.\\
\textit{Level 0 - Field Level:} Includes the physical process to be implemented.\\
\textit{Level 1 - Control Level:} Includes field controllers and devices (PLC, RTU) implemented for controlling the physical process.\\
\textit{Level 2 - Supervisory Level:} Includes the supervisory level controllers and devices (SCADA, HMI) implemented for controlling and monitoring of the field level controllers.\\
\textit{Level 3 - Operations/Monitoring Level:} Includes the components implemented for information sharing between the IT and OT layer (such as Workstation, HMI, DMZ).\\
\textit{Level 4 - Management/Enterprise Level:} Includes implementation of the management and planning resources at the enterprise level, such as resource management or risk assessment.

\textbf{Fidelity}, \textbf{Scale} and, \textbf{Cost:} These three inter-related domains represent the several properties of an ICS testbed, as described earlier in Section II-B. Evaluation of the fidelity domain for ICS-CTM2 framework is based on the testbed construction methodology adopted for implementation of three key components - Process, Protocol, and Infrastructure (PPI). Implementation of multiple process control scenarios and cyber-physical integration influence the scalability of a testbed. Cost of design is determined by factors such as physical components used, open-sourced or licensed products, training and development expenses.

\textbf{Application:} We have evaluated ICS testbeds based on the practical applications and use cases implemented \cite{Hahn2013}. The application scenarios have been divided into four categories in the order of increasing criticality: (1) Functionality Testing, (2) Education, (3) Cybersecurity Analysis and Research, and (4) Development of Standards .

\subsubsection{Maturity Indicator Level (MIL)}
ICS-CTM2 framework defines four Maturity Indicator Levels - MIL0 thru MIL3. A fifth maturity level (MIL4) is reserved for future use. The MILs helps to identify the degree of implementation in a domain. The MILs are independent to each domain, and to earn a MIL in a given domain, a testbed must satisfy all the evaluation criterion specified in that level. For instance, if a testbed satisfies all criteria in MIL1 and MIL2, but not in MIL3, then the testbed would be assigned a level of MIL2 in that domain.

\begin{figure*}
    \centering
    \includegraphics[width=\textwidth]{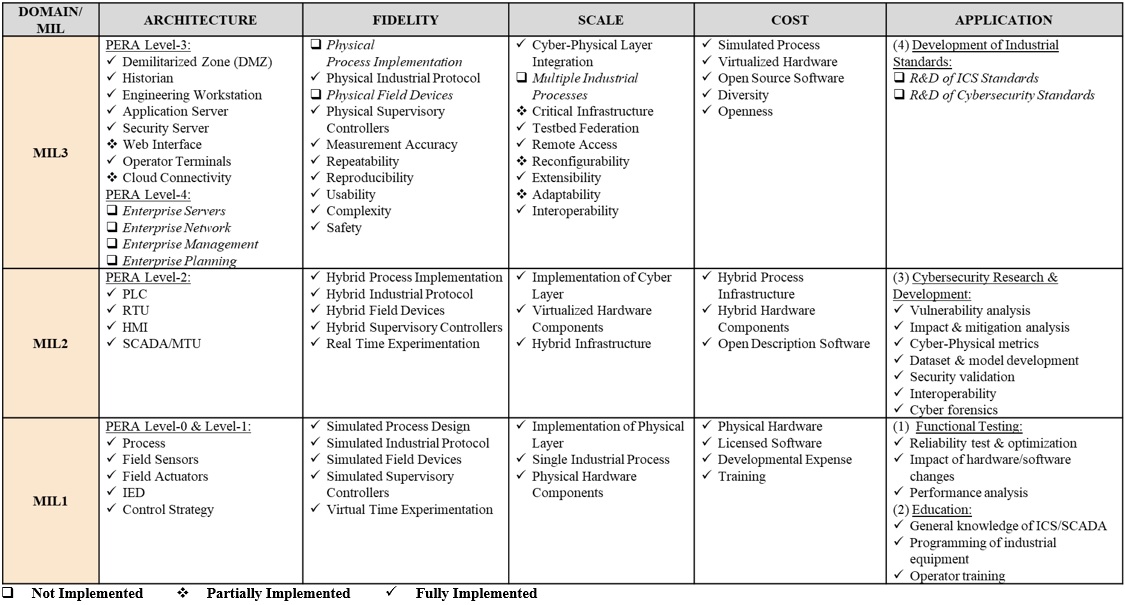}
    \caption{Evaluation Criterion considered within the ICS-CTM2 framework domains (case study performed for PowerCyber Testbed \cite{Ashok2016}).}
    \label{fig:ICS-M2}
    \vspace{-0.5cm}
\end{figure*}

\subsubsection{Diagnosis and Assessment:} For a testbed, each domain is assessed depending on the evaluation criterion and a MIL is assigned. First, we analyze a testbed with respect to each of the five domains. Analyzing a domain involves assessment of each evaluation criterion within that domain. Once all criteria are evaluated, a MIL rating is assigned to that domain. Similarly, all remaining domains are assessed. Finally, the ICS-CTM2 diagnosis results are visualized though Radar Analysis and Ring Analysis.

\textbf{Radar Analysis:}  The radar analysis visualizes the MIL ratings obtained by a testbed in each of the five domains. Radar analysis can be performed for a single as well as multiple ICS testbeds, and the results can be aggregated and displayed together for better comparison. All domains are arranged in a radial chart and the respective maturity ratings are recorded after implementation of ICS-CTM2 framework. Figure 3 and Figure 4 demonstrates the Radar Analysis.

\textbf{Ring Analysis:} The ring analysis provides a more in-depth assessment by illustrating the implementation degree of each evaluation criterion for all the five domains of the ICS-CTM2 framework. Ring analysis is a higher level evaluation, hence it must be performed for each testbed individually. As such, results of different testbeds cannot be combined for ring analysis. Figure 6 demonstrates the Ring Analysis.

\begin{figure*}
    \centering
    \includegraphics[width=\textwidth]{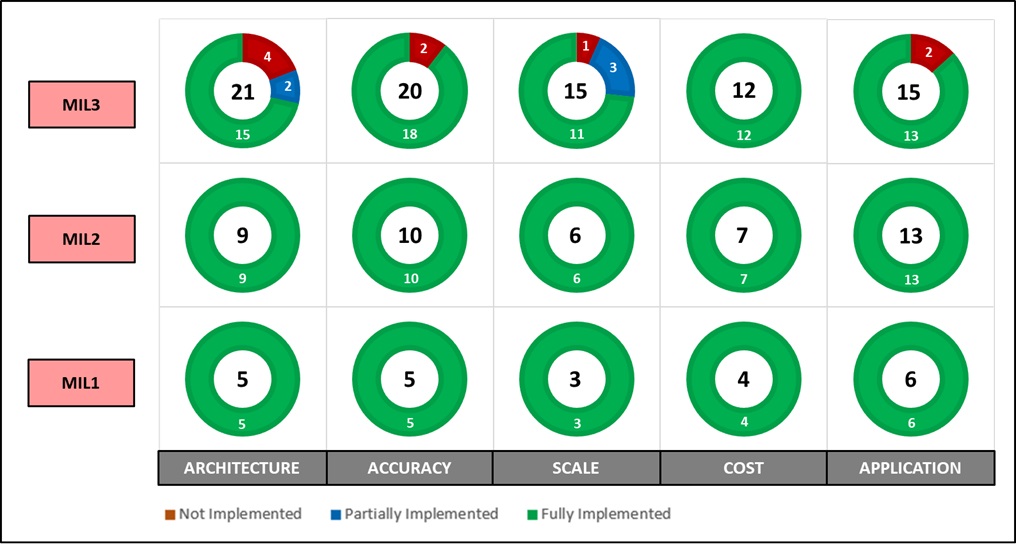}
    \caption{Case Study: Ring Analysis of Evaluation Criterion for all five domains (case study performed for PowerCyber Testbed \cite{Ashok2016}).}
    \label{fig:MIL}
    \vspace{-0.5cm}
\end{figure*}

\section{Case Study Analysis}
This section describes a case study implemented to analyze different testbeds, mentioned in Table I, using the ICS-CTM2 framework. In our experiment, we have considered one ICS testbed from 8 different industrial sectors, to demonstrate their differences. However, the ICS-CTM2 framework is not limited to this comparison only, and can be applied to any combination of ICS testbeds as deemed necessary.

Figure 2 provides the MIL ratings for the ICS-CTM2 domains after analyzing the selected testbeds. Figure 3 and Figure 4 represents the Radar Analysis comparing the performance of testbeds in each domain utilizing results demonstrated in Figure 2. The results of the selected testbeds have been divided into two plots for ease of illustration. As shown in the figures, from the Radar Analysis the researchers can easily identify the maturity of the testbed in each domain.

Figure 6 demonstrates the Ring Analysis for the ICS-CTM2 framework. We have performed this evaluation only for the PowerCyber Testbed \cite{Ashok2016} at Iowa State University. The evaluation criterion for each domain have been specified in Figure 5. The number specified in the inner ring indicates the total evaluation criterion considered at that maturity level. The number specified in the outer ring summarizes the total evaluation criterion satisfied or implemented (either fully, partially, or not implemented). For example, in Figure 6 Ring Analysis, the domain \textit{Architecture} has 5 evaluation criterion for MIL1; 4 criterion are added in the next level making the total 9 for MIL2, and 12 criterion are added in the next level making the total 21 for MIL3. The testbed satisfies all criterion for MIL1 and MIL2, but not for MIL3. Therefore, the testbed achieves an overall rating of MIL2 in this domain.

\section{Conclusion and Future Work}
The need for a standardized method to perform comparative assessment of ICS Testbeds is quite compelling. The design goal of the ICS-CTM2 framework, described in this paper, is to enable researchers to perform a self-evaluation of ICS Testbeds with respect to their design specifications and characteristic attributes, to determine their strengths and limitations. The ICS-CTM2 framework enables ICS testbeds to be analyzed on the basis of standardized domains which serve as benchmarks. Each evaluation criterion within the domains can be assessed to gauge the maturity in the design of a testbed. This paper also provides a case study analysis performed on 8 ICS testbeds to demonstrate the functionality of ICS-CTM2 framework and visualization of results.

\textbf{Future Work:} The implementation methodology and the framework domains described in this paper is being actively refined, to enhance the assessment parameters. The future work includes: (1) a comparative assessment of a larger number of ICS testbeds, applying ICS-CTM2 framework, based on the data about testbed specifications collected through scientific survey, (2) the proposed ICS-CTM2 model can be further refined by updating or expanding the domains, evaluation criterion, or the analysis method.

\section{Acknowledgement}
This work is funded in part by the NSF CPS grant ECCS 1739969.

\vspace{-0.2cm}
\bibliographystyle{ieeetr}
\bibliography{Bibtex}

\end{document}